\documentclass[conference]{IEEEtran}

\usepackage{multicol}
\usepackage{url}
\begin{document}
\title{Short Literature Review for a General Player Model Based on Behavlets}

\author{\IEEEauthorblockN{Benjamin Ultan Cowley}
\IEEEauthorblockA{BrainWork Research Centre\\
Finnish Institute of Occupational Health\\
POBox 40, Helsinki 00250, Finland\\
\\
Cognitive Brain Research Group\\
University of Helsinki, Finland\\
Email: ben.cowley@helsinki.fi}
\and
\IEEEauthorblockN{Darryl Charles}
\IEEEauthorblockA{School of Computing \& Information Engineering\\
University of Ulster\\
Northern Ireland\\
Email: dk.charles@ulster.ac.uk}
}

\maketitle

% As a general rule, do not put math, special symbols or citations in the abstract
\begin{abstract}
We present the first in a series of three academic essays which deal with the question of how to build a generalized player model. We begin with a proposition: a general model of players requires parameters for the subjective experience of play, including at least: player psychology, game structure, and actions of play. Based on this proposition, we pose three linked research questions, which make incomplete progress toward a generalised player model: 
\textsf{RQ1} \textit{what is a necessary and sufficient foundation to a general player model?}; 
\textsf{RQ2} \textit{can such a foundation improve performance of a computational intelligence-based player model?}; and 
\textsf{RQ3} \textit{can such a player model improve efficacy of adaptive artificial intelligence in games?} 
We set out the arguments behind these research questions in each of the three essays, presented as three preprints. 
The first essay, in this preprint, reviews the literature for the core foundations for a general player model. We then propose a plan for future work to systematically extend the review and thus provide an empirical answer to \textsf{RQ1} above. This work will directly support the proposed approach to address \textsf{RQ2} and \textsf{RQ3} above.
This review was developed to support our 'Behavlets' approach to player modelling; therefore if citing this work, please use the relevant citation: 

Cowley, B., \& Charles, D. (2016). Behavlets: a Method for Practical Player Modelling using Psychology-Based Player Traits and Domain Specific Features. \textit{User Modeling and User-Adapted Interaction}, 26(2), 257-306.
\end{abstract}


\section{Introduction}
\label{intro}
Understanding and modelling players, and the differences between them, can be considered an important milestone towards generalised human-level game playing AI. We define a player model as any system for identifying or interpreting player behaviours over sequences of game state variables from, e.g., game logs, or sensors. Such behaviours can be defined as patterns of player action that are consistent across repetitive game situations, where there are three main sources of influence which shape these reoccurring patterns: A) the human repertoire for engaging with tasks; B) the kinds of experience which game designers attempt to induce; C) the actions which are actually possible within a game. Thus we propose that a generalised player model must include parameters to describe subjective experience of play based on a foundation of established modelling tools, including: a) some way to describe psychology of behaviour; b) some language of general game design; and c) a model of actions in the context of a given game. This foundation should be established \textit{in addition} to any computational intelligence approach to derive a model from raw data.

This proposition implies the research question: \textsf{RQ1} \textit{what existing tools to describe player experience constitute a necessary and sufficient foundation for a generalised player model?}

A complete response to this question will require a systematic review and meta-analysis of the existing work in each of the outlined foundational areas, in order to establish the most useful and/or validated options for describing each area in a model. These types of review can provide replicable and quantified answers to well-defined research questions, so long as the relevant literature is mature and well-indexed. A short, non-systematic ‘scoping review’ can be used as a starting point; it aims to approximately describe the state of art and establish reference works, search terms and keywords to help build methodological search queries.

The aim of this preprint is to provide such a scoping review. The method of review was to gather sources on-demand, in support of our own work on player modelling, see \cite{Cowley2016behavlet}. Thus the review gives a good basic indication of the prior work that contributes to each of the foundation areas a) - c) above. Following the review, in section~\ref{disc} we describe a scheme for future work to systematise the review into a meta-analysis, which can give a quantified estimate of the efficacy of each foundational element that has been evaluated. The outcome of such future work will provide a comprehensive answer to \textsf{RQ1}, and also support research into further questions:

\textsf{RQ2} \textit{can such a foundation improve algorithmic performance of the computational intelligence required for a real-time player model?} (addressed in the second preprint \cite{Cowley2016prep2}); and 
\textsf{RQ3} \textit{can such a player model improve efficacy and viability of the artificial intelligence required to power games which adapt to their players?} (addressed in the third preprint \cite{Cowley2016prep3})

\section{Literature Review}
\label{litrev}
Two areas of research which can support player modelling but are under-exploited are game decomposition and player psychology. In Section~\ref{psych} we outline background research on player psychology. In Section~\ref{structure} we examine several key approaches for describing the composition of games. Section~\ref{experience} briefly describes some combined approaches which aim to create a framework of play experience. Finally Section~\ref{models} discusses some notable recent player models.

\subsection{Personality and Play}
\label{psych}
An important and often neglected aspect of player modelling is the influence of the player's personality on gameplay.

\cite{Koster2005} proposed that learning is fundamental to a game experience, supported by evidence from educational and comparative psychology \cite{Gee2003,Groos1898}; this theory indicates that models for describing ordinary differences in learning ability would also serve to describe differences in game playing experience. For example, individuals who are particularly good at analytical maths might excel when game mechanics call for abstract reasoning skills. In one such application, \cite{Acuna2010} attempted to model the behaviour of expert players of a game based on the Euclidean Travelling Salesman Problem, in order to show that such modelling could find novel solutions to NP-hard problems. Learning style-based models are a potential avenue for future complementary research in player modelling.

\cite{Salen2004a}'s implicit rules convey an additional aspect of games related to the Magic Circle concept \cite{Huizinga1949}; that is, the core experience of games represent an experiential space apart from normal life, which may be thought of as a kind of informal social contract where the normal rules of behaviour do not apply in the same way. This has additional implications, because the standard influences on an individual that arise from the interaction of personality and environment are weakened; within the novel space of a game, players may assume a play personality quite different from their own. For example, a player may be more aggressive than they would be in real life. Several player typologies have been proposed, beginning with Bartle's types for online multiplayer games \cite{Bartle1996}. However, players may still be subject to the influence of their basic underlying personal temperament type; it is generally difficult for people to adopt a persona that is fundamentally unaffected by their core personality type – it requires some ability to act. Thus the general personality models, such as temperament theory, can also be used.

Temperament theory describes modes of operation of human personality, how we act, react and interact \cite{Berens2006}. Two particular temperament theories \cite{Berens2006,Keirsey1984} are related to the popular Myers-Briggs Type Indicator (MBTI) \cite{Ludford2003} via a similar theoretical foundation, which in general proposes four categories that describe interrelated needs, values, talents and behaviours. For example, \cite{Keirsey1984} proposes an extension of Plato's classic types, Artisan, Guardian, Idealist, and Rational, with two categories and roles per type such that the whole model has 16 subtypes: correlating with the 16 types of Myers-Briggs. \cite{Berens2006} proposes four archetypes which are related to four skill sets: Logistical, Tactical, Strategic, and Diplomatic. Although the theory is not used as widely as trait models such as the 'Big 5' \cite{McCrae1992}, there is reason to believe that the two approaches are not strictly incompatible \cite{McCrae1989}, and types are a more useful approximation than traits when used in this context as a domain-shaping step. Temperament theory is a long established field of research, and as an approach that seeks to understand core motivation it provides a solid theoretical basis and useful skill-preference model in a game context. Temperament theory has been used before in human-computer interaction systems' research, and has served as an influence in a player typology called 'Brainhex' \cite{Bateman2011}. Each skill set is associated with preferred behaviour. 

Demographic Game Design (DGD) \cite{Bateman2005a} provides a useful example of a player typology as it is concise and represents a core set of types which tend to crop up across many typologies. DGD takes temperament theory, along with the MBTI, as its basis in describing player types and their associated game play-preferences, and has four types. As usual in typologies, players will belong to each of the types to a greater or lesser degree, because type membership is non-exclusive. The types can be briefly described as follows:
\begin{itemize}
	\item Conqueror: Competitive, win-at-all-costs. Players of this type are goal-oriented and enjoy feeling dominant in the game or in social circles set around the game.

	\item Manager: Logistical, plays to develop mastery. Such players are process-oriented and will replay completed games if they can use their new mastery to unearth novelty at deeper levels of detail.

	\item Wanderer: Desires new and fun experiences. Less challenge-oriented than the above types, these players primarily seek constant, undemanding and novel enjoyment.

	\item Participant: Enjoys social (living-room) play, or involvement in an alternate world.
\end{itemize}

Table~\ref{table:psych} is adapted from the work of Stewart \cite{Stewart2011}; it illustrates how richly the temperament theory can be mapped to existing game player models, thus supporting the general validity of our approach.

\begin{table*}[ht]
	\centering
	\caption{Mapping of Keirsey's four temperament types \cite{Keirsey1984} to a subset of four-quadrant game player type models from the literature \cite{Bartle1996,Caillois1961,Lazzaro2008,Hunicke2004a}; and \cite{Stewart2011}'s own conception of associated motivation, problem-solving style, and overall goals (adapted from \cite{Stewart2011}).}
	\label{table:psych}
	\begin{tabular}{llllllll}
		\textbf{Keirsey} & \textbf{Bartle} & \textbf{Caillois} & \textbf{Lazzaro} & \textbf{MDA+} & \textbf{Motivation} & \textbf{Problem-solving} & \textbf{Goal} \\
		~\cite{Keirsey1984} & ~\cite{Bartle1996} & ~\cite{Caillois1961} & ~\cite{Lazzaro2008} & ~\cite{Hunicke2004a} & & \textbf{ style} &\\
		\hline
		Artisan (tactical)    & Killer     & ilinx    & serious fun        & {[}kinetics{]} & Power (manipulative sensation)      & Performance           & Do           \\
		Guardian (logistical) & Achiever   & ag{\^o}n & hard fun, 'fiero'  & Mechanics      & Security (competitive accumulation) & Persistence           & Have         \\
		Rational (strategic)  & Explorer   & mimesis  & easy fun           & Dynamics       & Knowledge (logical rule-discovery)  & Perception            & Know         \\
		Idealist (diplomatic) & Socialiser & alea     & people fun         & Aesthetics     & Identity (emotional relationships)  & Persuasion            & Become      
	\end{tabular}
\end{table*}

\subsection{Game Structure}
\label{structure}
To unambiguously interpret player activity, it is necessary to decompose a game into its constituent parts. This is a non-trivial task due to the difficulty of defining games in the general case; as \cite[pp.27]{Wittgenstein1953} said on the commonality of features within games "you will not see something that is common to all, but similarities, relationships, and a whole series of them at that". We can examine a game from many perspectives: viewed as a formal system built of entities and rules, a game is mechanistic; viewed as an interactive system exhibiting emergent properties, a game is dynamic; viewed as an emotional experience, a game is aesthetic. These mechanics, dynamics, and aesthetics form the three perspectives in LeBlanc's MDA (Mechanics Dynamics Aesthetics) model \cite{Hunicke2004a}. In the seminal work of \cite{Salen2004a}, seventeen separate perspectives are used to examine games as systems ('Rules schema'), experiences ('Play schema'), and cultural artefacts ('Culture schema'). \cite{Salen2004a} also contribute the concept of Constituative (sic), Operative, and Implicit rule sets.

Several authors have considered gameplay patterns from a game design standpoint. For example, \cite{Koster2005} focuses on the inherent fascination that people have for patterns as a motivation to play, outlines how this relates to our desire to learn, and discusses the relationship of learning to the experience of fun. The basic concept is that a game contains patterns of activity that are initially unfamiliar to a player, but progress through the game corresponds to an increased understanding of these patterns, and a concurrent increase in skill. Ideas on game play patterns can be related to information theory and in particular entropy and uncertainty \cite{Salen2004a}. For example, \cite{Costikyan2013} postulates that uncertainty is a core ingredient of games; by this postulate, a player must be unsure of the outcome of a game to maintain interest. \cite{LeBlanc2006} has also discussed the role of uncertainty in the context of Formal Abstract Design Tools \cite{Church2006} to enhance dramatic tension within games. Patterns of play are also related to game design patterns. Game design patterns are readily identifiable objects of play which form the common core of many different games. The pattern consumption that \cite{Koster2005} discusses relates to the unique composition of game design patterns (and other novel game mechanics) for each individual game, which generates a unique distribution of information across the space of play and therefore represents a unique experience of learning as players sample the possibility space and estimate the distributions.

Several attempts have been made to develop a more formal approach to game analysis, including early proposals from game designers \cite{Church2006,Hunicke2004a}. They suggest that more effective analysis of games requires a descriptive grammar of play to underpin a common practical game design language. One approach is to break a game down into its most basic parts, e.g. atoms \cite{Cousins2006} or to build a practical collection of game design patterns, e.g. the 400 Project \cite{Barwood2006,RouseIII2015}. \cite{Bethke2003} used the Unified Modelling Language (UML) to define game elements, giving an example for Pac-Man, among others.

In Chapter 2 of \cite{Brathwaite2009}, the authors lay out the elements of a game, including a neat definition of game space as the embodiment of the game, game state as all game variables, and game view as what a player can access at a given moment. \cite[p.45]{Jarvinen2009} produced a comprehensive "theory about the parts that games are made of." He defines some useful concepts, such as components: "objects that the player is able to manipulate and possess in the course of the game" \cite[p.63]{Jarvinen2009}, broken down among components possessed and controlled by oneself, by others, or by the game system. This object-oriented view relates well to category theory \cite{Walters1991}, a method of formal modelling often used for proving computational systems and employed to define games by \cite{Grunvogel2005}.

Game designers have consistently used game design patterns over the years, either intentionally or intuitively, and so this is a natural method to use. The game ontology project (GOP) \cite{Zagal2005} is a simple structure of four categories and one hierarchical level that captures the important structural elements of games and relationships between them. \cite{Schell2008} provides an elegant analytic framework in his 'Book of Lenses'. Bj{\"o}rk and Holopainen \cite{Bjork2005} have completed some of the most comprehensive research on a complete framework for describing games in terms of game design patterns. Bj{\"o}rk and Holopainen describe a game design pattern as "\textit{semi-formalised interdependent descriptions of commonly reoccurring parts of the design of a game that concern gameplay}" \cite{Bjork2006}.

Bj{\"o}rk and Holopainen take a two stage approach in their method. They first describe a component framework where invariant aspects of gameplay can be mapped to games. Their framework has the following top-level game components: \textbf{Holistic}, \textbf{Boundary}, \textbf{Temporal}, and \textbf{Structural}. Holistic components define a game's unique character, including game instance (e.g. play location, time expended, player experience) and session (i.e. particular gaming activity within an instance). Boundary components define the purpose of a game, including the limits placed on a player as they attempt to achieve goals or overcome challenges. Rules that define what a player can or cannot do may be endogenous or exogenous, in that they may be explicitly coded into game mechanics or implicitly agreed between players and designers. Temporal components of games define the game's series of actions and events. Events give rise to a change of player and/or world states which require a player to make decisions and consequently perform actions to enhance or advance the current player and/or world state. Structural components define tangible game elements, such as user interface, game world, etc.

The second part of Bj{\"o}rk and Holopainen's approach is to define commonly reoccurring design patterns which can fit into the component framework. Components provide an abstraction of a game, but game design patterns describe how specific components interact to provide gameplay. The patterns are established using a designer-like approach by collecting and describing different events and components from the game then reflecting on how each one relates to the game playing experience. The result is an ever-growing list of reoccurring patterns in games, which allow us to describe the design of games and gameplay in a comprehensive manner. 

In their published collection, \cite{Bjork2005} describe over 200 patterns found repeatedly across different games. An example is the \textit{Aim \& Shoot} pattern, very common in many game types, not only first person shooters. This pattern involves dexterity-based action where one needs to pinpoint a target from a simulated space in real time and then initiate shooting \cite[p.150-153]{Bjork2005}. A different example which demonstrates how design patterns can capture non-sequential elements is the pattern \textit{Perfect Information}. Games utilizing this pattern never hide or keep secret any elements of the game from the player, nor depend on random input, for example Chess or Go \cite[p.128-130]{Bjork2005}. The patterns can be grouped to reflect patterns of similar qualities and scope. Bj{\"o}rk and Holopainen cover patterns in eleven major groups, including \textit{resource management}, \textit{social interaction}, \textit{game session}, and \textit{replay value}. The game design template used by Bj{\"o}rk and Holopainen contains seven categories which cover descriptions of the pattern, how the pattern may be used, consequences or limitations, and relationships to other patterns (e.g. instantiating, modulating, or conflicting). \cite{Lankoski2011} have also shown how design patterns can be theoretically derived, extending the value of the system beyond the limits imposed by the previous requirement of manual analysis. Bj{\"o}rk and Holopainen's game design patterns support analysis of what the player actually does in the game, and provide a common vocabulary and well-formed structure.

\subsection{Play Experience Frameworks}
\label{experience}
Game decomposition and player psychology models must be combined for maximum efficacy. This requires going beyond consideration of separate perspectives, e.g. MDA, to the definition of a framework that links perspectives with a scientifically plausible theory. For example, such a framework should account not only for the learning that occurs throughout a game, but also the emotions of players and their impact on play.

The User-System-Experience (USE) model was an early attempt at such a framework proposed by the authors \cite{Cowley2006a,Cowley2008}. In this framework the most novel aspect was the formulation of the pleasurable and autotelic nature of games in terms of Flow theory and the neurobiology of information processing and learning. However the specification of games themselves was lacking in detail. Another method, similar to what we propose here, has been demonstrated for the domain of educational games \cite{Bedek2011,Cowley2012}; here players were modelled in terms of the competences they show in the pedagogical domain. Competence model validity is improved by theory-driven expert design of features, because the psychological underpinnings of real-world competences are defined in the literature. This method works hierarchically from a description of a competence such as communication; through sub-competences, such as verbal and non-verbal communication; to a set of behavioural indicators based on empirical findings, e.g. non-verbal communication is indicated by listening, body language, proxemics (personal space); to contextual performance indicators represented by a formula defined over in-game variables, e.g. for proxemics there is position coordinates, viewport, etc. Such complexity shows that domain expertise is often needed in creating serious games \cite{Marchiori2012}, but less has been done to systematise that knowledge for player modelling, although techniques from intelligent tutoring have long served as inspiration for profiling in entertainment games \cite{Beal2002}. \cite{Herbert2014} took a similar approach to build a model of a gamified learner using gamification types based on combined psychology and gamer types, also tracking behaviour within a learning virtual world to investigate and refine the model.

\cite[p.99-247]{Jarvinen2009} proposed a model of player experience that builds on his game decomposition theory. Central to this model are two concepts: that game experiences are composed of sequences of emotions; and that game elements embody conditions that elicit emotions. It is also important that emotions are part of the cognitive game; the player is seen as predictive of her own and other players' emotions, which forms a reciprocity between emotion elicitation and active play. \cite{Gmytrasiewicz2000} also attempted a formalism of 'synthetic' emotions using Decision Theory, to be used for player modelling or for communication of AI agent states to the player.

Formal specification of play is a tool which can help to include psychological theory in player models. \cite{VonNeuman1944} defined the classic game theory, which gives useful tools to analyse player behaviour: assuming that players are rational agents with definable utilities for action. More generally, formal methods such as category theory \cite{Walters1991}, enable specification and verification of the objects and actions of the play space, and thus support rigorous testing of system coherence. While it is not a substitute for play testing, there are many advantages in testing algorithms and functions. Formal methods of category theory were applied to game specification in \cite{Grunvogel2005}, which seems mainly aimed at illustrating the pros and cons of formal modelling for games, and draws heavily on the more complete work of \cite{Tabuada2004} on modelling abstract control systems. Thus while not fully developed there is potential here for further work. In \cite{Grunvogel2005}'s abstract specification, a game “consists of objects which change their state during the play, where the evolution of their state is governed by rules and influenced by the players or other objects”. This approach is flexible, but the complexity of the domain poses a large problem for this method. The author agrees: “describing a game with this formalism seems to be a cumbersome task”. This is as a result of the attempt to encapsulate all aspects of all possible games in one system; any such attempts will be either unwieldy or insufficiently descriptive.

Attempts have been made to codify game mechanics. \cite{Sicart2008} uses the object oriented (OO) programming paradigm to define game mechanics as “methods invoked by agents”. This offers to a modelling tool an OO-like process description schematic, which would allow prediction of the outcomes of player actions. If rationality and complete information are assumed, then the player's preferred outcome will be foreknown. Such assumptions are not usually considered tenable in player modelling, however.

\cite{Breining2011} developed a formal modelling tool-set to analyse player behaviour by action sequence mining. The method finds all action sequences and their frequency in a game log, representing common sequences as features, which are ranked by mutual information with the class variable for feature selection. This is a useful tool for characterising players such that two players can be distinguished by their game logs alone.

\subsection{Player Modelling}
\label{models}
Player models have often been constructed using computational intelligence methods, to extract statistical features of play and cluster or classify players. For general reading on machine learning in games and player modelling, see e.g. \cite{Bakkes2012,Galway2009}. Here we will only touch on some examples which illustrate particular difficulties of computational player modelling.

Computational intelligence for player modelling faces a number of subtle and non-trivial difficulties. Automated feature extraction in the domain of games using only computational means depends on the particular game mechanics to be learned, e.g. number of players, stochastic game mechanics, etc.. Classically the field has focused on board games, from \cite{Samuel1959}'s checkers player, through \cite{Tesauro1995}'s Temporal Difference Learning in Backgammon (see also \cite{Ghory2004}) up to the recent deep-learning AlphaGo program which beat a world-class Go champion \cite{Silver2016}. Despite the success of the latter work, feature extraction for generalised game play remains challenging - \cite{Mnih2015} has demonstrated a solution using deep-learning, but this also demonstrates the limitations of the state of the art, because it was successful only for very simple games with short time horizons.

Other work has also focused on extracting maximum information about player behaviour from simple metrics. \cite{Drachen2009} analysed a large data set from of complete games of \textit{Tomb Raider: Underworld} (Eidos, 2008), using unsupervised artificial neural networks (ANNs) to cluster and visualise a set of six simple game metrics. The clusters enabled classification of players into four ad hoc personas (Veterans, Solvers, Pacifists and Runners). These types were labelled by the manual process of examining the typical game play behaviour associated with each cluster. While this approach is a straightforward way to obtain a player profile, the method gives no guarantee to converge to interpretable clusters. In another approach, \cite{Holmgard2014} designed agents they call “procedural personas”, which can then be compared to human play and evolved to emulate it. In both these works, the outcomes do not necessary link well to established theories of personality or temperament.

\cite{Houlette2004} has looked at methods to automatically model players based on factorization of their game-play attributes, for example in an FPS, recording factors (such as ratio of shots-fired to shots-on-target) that comprise the player's accuracy attribute. This algorithm can be directed, as in \cite{Houlette2004}, or be self-organizing, as in \cite{Charles2005}. The former may be more computationally efficient but is less generic and relies on expert input to define factors and attributes. The latter method would use unsupervised neural networks to build correlations from raw data sets, and so create factors that are not predefined by algorithm designers but arise naturally from play.

A popular approach to adapting a game is some form of Dynamic Difficulty Adjustment (DDA) – for example, by altering the number of power-ups in a game or by making non-player characters more or less co-operative or competitive.  Several games have attempted to implement DDA systems, e.g. \textit{Max Payne} (3D Realms, 2001) and \textit{Prey} (3D Realms, 2006). Difficulty is indeed a key factor to tune for an optimal player experience, and DDA has resulted in some interesting approaches, e.g. \cite{Chanel2011}. However, there are many other factors which impact player engagement. Both player-selected difficulty settings and DDA typically account for variation in capability but not in player type. Features for DDA are not designed to enable a deduction about player psychology but rather to tune the 'game-challenge' utility function, whereas a general model should provide insight into different facets of player behaviour, for example the information processing 'style' of a player. There is thus scope for much richer player models.

In previous work, we have partly addressed this challenge with the \textit{Behavlets} system for creating player-modelling features linked to valid psychological theory \cite{Cowley2016behavlet}. The Behavlet method draws together three lines of work, released in several of the first author's existing publications, so to illustrate the novelty of this paper it is appropriate to describe these foundations. In the first line, \cite[p.77-90]{Cowley2009} describes how we devised a na{\"i}ve concept of behaviour traits and the 'constraint harness' approach (defined in \cite{Cowley2016behavlet}) to derive features descriptive of player type. These concepts were applied in \cite{Cowley2012a} to show how analysis of Decision Tree player models could benefit from richer features. In the second line, the core Behavlet ideas, inspired by the architectural patterns of \cite{Alexander1990}, were developed in the context of serious games development for behaviour change \cite{Cowley2011b}. Behavlets in that work described patterns of real-world energy-use behaviour that could be leveraged by simulation game designers. Finally, the third line arose from \cite{Cowley2014a}, in which game design patterns were related to clusters of game events learned from time series of played games using the FP-growth algorithm. That work showed how disparate methods could be integrated, but itself lacked a means to describe the player's actions in psychological terms. The Behavlets represents a next step, but it remains incompletely evaluated and further work should aim for more integration with the approach from \cite{Cowley2014a}.

\section{Discussion}
\label{disc}
The purpose of this review is to 'scope out' the background for the argument that generalized player model should build on the foundation of a broad and deep description of the game, including: psychology of play; structure of games; and frameworks of experience. Such general models would go beyond the 'standard' approach of simple features constructed from in-game variables, and build rich features with psychological and game design theoretic foundations.

An important step in future work will be to systematically expand this review into meta-analyses of the field, for example to evaluate the efficacy of various existing player models. With an empirical analysis it will be more feasible to assess which elements make the best models successful, and determine how a general approach can benefit from and contribute to good modelling.

The review process starts by stating research questions of interest for each area, for example, asking for psychology of play "what is the accuracy of player classifiers based on Bartle's typology?" \cite{Bartle1996} A defined search strategy enables the researcher to pose the necessary and sufficient research questions, e.g. by finding all published player typologies. Next, systematic review obtains all relevant literature from indexed databases, by creating a replicable methodology for the review process, founded on defined search and filter procedures. For example, for each player typology the citing articles are obtained from e.g. ACM's Digital Library. These articles are then filtered to those which propose or evaluate a player model. In the domains of interest it may be necessary to relax the usual requirements for search replicability, because the relevant literature (published in, e.g. conference workshops) is often not indexed as thoroughly as the medical research targeted for traditional systematic review.

The systematic reviews will provide a quantitative sampling of the literature, giving a distribution of values for each research question. Meta-analysis follows to extract quantitative answers, such as general classifier goodness when using Bartle's typology. The information provided by the review process will then help to explore \textsf{RQ2}, creating the foundation to enhance an existing computational intelligence algorithm.

\section{Conclusion}
\label{conc}
In this review we have described selected works from three areas that support player modelling. We also reviewed a small part of the player modelling literature where it helps to illustrate some of the challenges in the field. The main aim is to provide a snapshot of the player modelling state of art, and propose a strategy for validating that snapshot against the 'ground truth' through systematic review and meta-analysis.

\section*{Acknowledgements}
Partly supported by the Finnish Funding Agency for Innovation, project Re:Know \#5159/31/2014


\bibliographystyle{abbrv}
\bibliography{Cowley_Behavlet_preprint_1_bib}

%\printbibheading
%\printbibliography[]

\end{document}